\PassOptionsToPackage{table,xcdraw}{xcolor}

\documentclass[screen,authorversion]{acmart}

\AtBeginDocument{%
  }

\copyrightyear{2023}
\acmYear{2023}
\setcopyright{rightsretained}
\acmConference[LAK 2023]{LAK23: 13th International Learning Analytics and Knowledge Conference}{March 13--17, 2023}{Arlington, TX, USA}
\acmBooktitle{LAK23: 13th International Learning Analytics and Knowledge Conference (LAK 2023), March 13--17, 2023, Arlington, TX, USA}
\acmDOI{10.1145/3576050.3576081}
\acmISBN{978-1-4503-9865-7/23/03}

\begin{document}

\title{Insights into undergraduate pathways using course load analytics}

\author{Conrad Borchers}
\affiliation{
  \institution{Carnegie Mellon University}
  \streetaddress{}
  \city{}
  \country{USA}
}
\email{cborcher@cs.cmu.edu}

\author{Zachary A. Pardos}
\affiliation{
  \institution{University of California, Berkeley}
  \streetaddress{}
  \city{}
  \country{USA}
}
\email{pardos@berkeley.edu}

\renewcommand{\shortauthors}{Conrad Borchers and Zachary A. Pardos}

\begin{abstract}
Course load analytics (CLA) inferred from LMS and enrollment features can offer a more accurate representation of course workload to students than credit hours and potentially aid in their course selection decisions. In this study, we produce and evaluate the first machine-learned predictions of student course load ratings and generalize our model to the full 10,000 course catalog of a large public university. We then retrospectively analyze longitudinal differences in the semester load of student course selections throughout their degree. CLA by semester shows that a student's first semester at the university is among their highest load semesters, as opposed to a credit hour-based analysis, which would indicate it is among their lowest. Investigating what role predicted course load may play in program retention, we find that students who maintain a semester load that is low as measured by credit hours but high as measured by CLA are more likely to leave their program of study. This discrepancy in course load is particularly pertinent in STEM and associated with high prerequisite courses. Our findings have implications for academic advising, institutional handling of the freshman experience, and student-facing analytics to help students better plan, anticipate, and prepare for their selected courses.
\end{abstract}



\maketitle

\section{Introduction}

Course pathways through higher education degree programs are part of a learning arch that unfolds as a function of a learner's resources, constraints, and decisions \cite{de2020learning}. Compared to K-12, US institutions of higher education, particularly four-year universities, give students a high amount of elective course choice. This choice comes with unique challenges that can inhibit their learning path, such as the choice to overload on credit hours causing early undergraduate dropout among older students with prior vocational training and completed degrees \cite{karimi2022causal}. Conversely, low enrollment levels have also been found to be associated with worse educational outcomes, potentially due to a lack of financial and academic support \cite{boumi2021quantifying}. These findings, though seemingly contradictory, suggest that semester workload may play an important role in explaining the complicated story of student success in higher education. However, recent work has found that credit hours is not a suitable proxy for course workload, as it captures only 6\% of the variance in student reported course load compared to 36\% captured by LMS features \cite{pardos2023credit}.

In this paper, we introduce course load analytics (CLA) as a machine learning approach to producing metrics about course workload relevant to student course selection. This work is the first to predict course load at scale, generalizing to over 10,000 courses at a large public institution and going beyond time load considerations by incorporating more holistic measures such as mental effort and psychological stress. Our findings suggest that the discrepancy between anticipated course load (i.e., as calculated by credit hours) and actual course load (i.e., as estimated by CLA) may be a significant factor in program stop-out. 

We define CLA as predictions or inferences of course-level workload based on institutional data (e.g., enrollment data) and student-level course attributes (e.g., average student GPA). Our approach is novel in two principal ways. First, we expand previous approaches to utilizing student workload for academic outcome prediction that have been largely reliant on time load and credit hours \cite{karimi2022causal} by adding students' course perceptions regarding mental effort and psychological stress. Second, we train a machine learning model that can be used for course advising and studying higher education pathways.

We find that our model verges on readiness for real-world deployment, improving accuracy by 22\% compared to an average baseline. This establishes a benchmark for course load prediction that future work may build on. We contribute a novel methodology to derive CLA via student surveys and facilitate reproducible procedures to study student pathways with regards to predicted workload across time. Compared to prior works on student ``at-risk'' prediction (cf. \cite{parker1999study,arnold2012course,jayaprakash2014early}) which focus on predicting drop-out for a \textit{student} with respect to a course, our approach focuses on estimating workload as a construct of a \textit{course}, independent of a particular student. We demonstrate the potential of CLA by drawing novel insights about undergraduate pathways via retrospective analysis of historical enrollments, including group comparisons. Our results highlight the role of course prerequisites, first-semester course load, and unexpected excess predicted course load compared to credit hour designation in the outcomes of degree stop-out and on-time graduation.

\section{Related work}

Credit hours historically emerged as a standardized measure of student time spent on courses necessitated by the introduction of the elective system in North America \cite{heffernan1973credibility, shedd2003history}. As such, credit hours alone do not convey the workload associated with a particular course regarding other dimensions (e.g., mental effort and psychological stress). Work psychology has suggested that workload is a multi-dimensional construct comprising time load, mental effort, and psychological stress \cite{reid1988subjective}. While mental effort can be described as the difficulty of learning complex cognitive tasks \cite{paas2010cognitive}, psychological stress has been measured as excess workload in the workplace via self-assessment \cite{ilies2010psychological}. Prior works on the relationship between higher education workload and outcomes have been largely operating on time-based workload metrics. While a recent study suggested that reducing academic workload defined in credit hours may reduce dropout risk \cite{karimi2022causal}, another study indicated that time spent on coursework increases before assignment deadlines \cite{ruiz2011assessing}. Still, we know that mental effort matters in student outcomes. Math and English preparedness, representations of student prerequisites when dealing with advanced subject matter, are significant correlates of STEM degree attainment in community college transfer students \cite{chen2013stem, zhang2022early}. Similarly, work on students' coping mechanisms when dealing with high workloads and resulting psychological stress is scarce. Prior work suggested that students may cut into their leisure to cope with high time load and avoid low grades \cite{huntington2021semester}, potentially to the detriment of their well-being \cite{smith2019student}. Still, there is a dearth of studies that have looked at the longitudinal effects of high workload on university students, with prior works focusing on the effects of procrastination \cite{waschle2014procrastination} and socio-emotional competencies \cite{parker2006emotional} on academic outcomes. 

\subsection{Course recommendations and course-level analytics}

One area in which learning analytics has made advancements to improve student-facing course selection regarding academic outcomes is course feedback and planning tools. We identify three main threads in that area.

The first thread relates to personalized student feedback and early warning systems based on student and course attributes. For example, the Course Signals System at Purdue University allowed instructors to provide course feedback to students based on learning analytics that predict successful course completion and foster retention \cite{arnold2012course}. Related work scrutinized the ability of predictive models to identify struggling students based on their academic attributes \cite{jayaprakash2014early}.

The second thread relates to course and enrollment analytics for academic planning and success. \citet{pardos2020designing} designed course recommendations to be novel or unexpected to students but still relevant to their interests using catalog and enrollment-based course representations. Related work leveraged attention models to curate course recommendations based on students' past enrollment history and future courses of interest across semesters \cite{shao2021degree}. Other studies employed co-enrollment sequences to predict student grades in higher education via graph convolutional networks \cite{hu2019academic} and hidden Markov models \cite{boumi2019application}. The CARTA project provided student-facing course statistics, such as historical grade distributions and aggregate ratings from course evaluations, to support student course selection decisions \cite{chaturapruek2021studying}.

The third thread relates to learning analytics to improve instructional and program design. At the course level, prior work has linked instructional design, particularly instructor activities, to student engagement in an LMS \cite{ma2015examining}. At the institutional level, course-level analytics may help craft design feedback loops. Prior work found time- and performance-related metrics gleaned from program compositions to enhance their curriculum \cite{ochoa2016simple}.

\subsection{Predictions of higher education degree attainment}

Recent advances in representation learning have been applied to learner interactions with educational platforms, including user-item interaction \cite{wang2021mooc}, the timing of learning sessions \cite{li2022dropout}, and clickstream data \cite{le2018communication,jin2021dropout} to improve dropout prediction in MOOCs. In higher education, program dropout may take various forms \cite{tinto1975dropout} and might be confounded with protected academic attributes \cite{yu2021should} as well as student-level beliefs \cite{dai2014changes} and traits \cite{parker2006emotional}.

Analytics that allow integrating data on these various factors contributing to higher education dropout with nascent machine learning methodologies may be crucial to improve our understanding of how student trajectories culminate in program stop-out and on-time graduation. While multiple studies have presented viable models of higher education dropout prediction \cite{nagy2018predicting, baranyi2020interpretable}, these models left room for improvement or, at least, exploration \cite{kumar2017literature}. Selected works in the context of distance education have made use of these opportunities in predictive modeling. One study in the context of distance education in community colleges took into account locus of control and source of financial assistance to predict dropout with close to 85\% accuracy \cite{parker1999study}. Later studies expanded and showed that satisfaction in e-learning is significantly related to dropout \cite{levy2007comparing}. Notably, early work using causal models found institutional commitment and routinization to be strong determinants of higher education dropout \cite{bean1980dropouts}.

Prior work on on-time graduation leveraged features encapsulating relationships between a student's traits, academic environment, and other contextual factors. A survey of over 4,500 undergraduates in the United States concluded that college environment and personal financial characteristics are strong determinants of time-to-degree \cite{letkiewicz2014path}. Few studies have aimed to leverage features representing these attributes to predict academic outcomes in higher education. \citet{herzog2006estimating} predicted on-time graduation with student demographic, academic, residential, and financial aid information. Similarly, \cite{luo2018diagnosing} improved predictions of on-time graduation via historical enrollment sequences in Integrative Biology.

To the best of our knowledge, there has been no systematic study on how workload characteristics of courses may contribute to stop-out and on-time graduation in higher education. This appears to be much needed, given that course prerequisites, which may impose a higher workload if not adequately satisfied, are frequently cited as a determinant of STEM achievement. For example, prerequisites are associated with first-year academic achievement in university mathematics \cite{kosiol2019mathematics}, and perceived math preparedness is associated with STEM persistence \cite{dika2016early}.

\subsection{The present study}

Our research questions revolve around the validity of our proposed CLA framework and include a two-step process for validation and exploration. We investigated the internal validity of our construct through predictive cross-validation based on our ground-truth survey data. We then applied predictions to unlabeled courses to seek insights into how course load plays into student pathways and academic outcomes. Our research questions are as follows:

\begin{itemize}
    \item[] RQ1: How accurately can we predict course load via LMS, enrollment, and course embedding features?

    \item[] RQ2: How does the semester course load of different types of students compare throughout their academic degree? Are there differences between STEM and non-STEM majors, transfer and non-transfer students?
    
     \item[] RQ3: How does a student's predicted semester course load interact with credit hour load with respect to degree stop-out and on-time graduation? 
\end{itemize}

\section{Datasets}

The unit of analysis for our course load prediction is individual courses. Surveyed student ratings of course load were used as the ground truth in our dataset, with aggregate ratings per course serving as the targets of our models. For model features, we concatenated (1) LMS features, (2) student and course academic attributes from enrollment data, and (3) pre-trained course2vec features representing co-enrollment sequences for each course in our sample on a per-semester basis. As we describe in the next subsection, we collected our training data in the Spring 2021 semester.

\subsection{Ground-truth survey ratings of course workload}
\label{sec:data:survey-data-collection}

Immediately following the end of the Spring 2021 semester at the University of California, Berkeley (UC Berkeley), we asked undergraduate students to rate the lecture-based courses they had taken that semester with regard to workload. Our survey sample included 596 course load ratings from 128 students. Drawing from the Subjective Workload Assessment Technique (SWAT) \citep{reid1988subjective}, we asked students to rate each course on time load, mental effort, and psychological stress with respect to workload magnitude and manageability, yielding two survey questions per construct. We measured all course load dimensions on five-point Likert scales, with their wording adapted to the higher education context as described in \citet{pardos2023credit}.\footnote{Time load was rated on a six-point Likert scale at data collection. We truncated the scale for standardization and the purpose of this study by treating a response of either a six or a five on the time load scale as a five, which included around 4\% of total responses.} The three course load subconstructs had satisfactory reliability based on correlations between magnitude and manageability ratings (between -0.56 and -0.79) and exhibited moderate to high intercorrelations (between 0.54 and 0.67). Therefore, we analyzed a fourth course load metric in this study: the average workload magnitude and manageability rating of a course across time load, mental effort, and psychological stress. As an example, if a student rated a course a 2, 3, and 4 on time load, mental effort, and psychological stress magnitude, then that student's combined course load magnitude rating for that course would be $\frac{2+3+4}{3} = 3$. Participants received a \$15 gift card for survey participation. As courses were rated an average of 1.80 times, we first averaged multiple ratings for each course. This ensured that courses with repeated entries in the training data would not bias the model weights and leak to the test set. This procedure yielded a training dataset of $N$ = 332 course workload ratings.

\subsection{LMS data}
\label{sec:method:lmsdata}

We engineered features expected to be associated with course workload from historical Canvas learning management system (LMS) data at UC Berkeley. Our data included complete LMS records of all courses taught at UC Berkeley between Spring 2017 and Spring 2021. The unit of analysis is time-stamped student or instructor interactions with the LMS. Our features related to assignments (13 features; e.g., number of assignments and the spread of their deadlines), submission comments (4 features; e.g., submission response rate and average size), and forum posts (14 features; e.g., average reply speed and frequency) made by students, teaching assistants, and instructors (Table \ref{tab:discrepancy-determinants}). We aggregated our features on a semester level, given potential distributional shifts across course offerings. Dealing with missing LMS data, we compare a $k$-means imputation to a control variable imputation strategy at cross-validation. For the control variable approach, missing values are kept at 0 while a binary control variable, for example, for general lack of missing LMS forum data, is included as a feature. We concatenated data based on all available Canvas courses associated with a particular course for a specific semester, given that part of the forum and assignment activity may be scattered across multiple Canvas courses. An overview of all employed LMS features, including reproducible code for Canvas feature engineering and extensive descriptions of their design is in \citet{pardos2023credit}.

\subsection{Enrollment data and academic outcomes}

We created course features from historical student course enrollments. These features included credit hour designation, course GPA, course grade standard deviation, \% and availability of non-letter grades, \% of pass or satisfactory among non-letter grades, and within-major and overall  GPA of enrolled students. In addition, we coded the STEM status of courses and their student composition based on academic departments at UC Berkeley, following STEM designations by the U.S. Department of Homeland Security.\footnote{\url{https://www.ice.gov/sites/default/files/documents/stem-list.pdf}} Finally, we computed the total number of prerequisites and the total and relative average of satisfied prerequisites by enrolled students in the observed and all past semesters. We defined degree stop-out as students having at least two consecutive semesters of absence and delayed graduation as students not graduating within eight semesters for non-transfer students and four semesters for transfer students.

\subsection{course2vec}

We used pre-trained 300-dimensional course vector representations (i.e., course2vec) from prior work as additional features for prediction \cite{pardos2020designing}. The unit of analysis of course2vec's training data is student-level enrollment sequences across multiple semesters. These course embeddings were trained analogously to natural language via a skip-gram model, with courses being analogous to words and enrollment sequences being analogous to sentences. The method embeds courses into a semantic space with lexical relations between courses similar to those found in language. For example, validation work of the vector representation correctly marked pre-specified course analogies such as Math 1B is to Honors Math 1B as Physics 7B is to Honors Physics 7B \cite{pardos2020university}. For this study, the co-enrollment information encoded in these embeddings may help predict course load. In addition to the course2vec of the course, we also add as a feature the average course2vec vector of all of the course's prerequisites. For courses without any prerequisites, this feature was imputed to be the average of all course2vec vectors.

\section{Methods}

Our methods comprise three steps. First, we describe our machine learning models and their training procedure. Second, we define model performance evaluation metrics and the model selection process for course load prediction to historical data that can most generalize out-of-sample. Third, we describe analyses conducted on historical enrollment data to investigate CLA regarding academic outcomes (i.e., one-time graduation and degree stop-out).

\subsection{Model training}

Out of our $N$ = 332 course load ratings, we kept 15\% of observations for a true holdout test set to evaluate model performance after cross-validation-based model selection. We ran five-fold cross-validation on the remaining 85\% of the data to pre-select models for test set evaluation such that, in effect, we used a little under 70\% ($0.8*0.85=0.68$) of the data for training in each phase of the cross-validation. At this step, we employed random search ($N$ = 25 draws) from uniform distributions of hyperparameters for all of our models. The number of hyperparameter searches was based on time constraints during training given the objective of 216 experiments completing in three weeks on a machine with 48 Intel Xeon CPU cores and 376 GB of system memory. The objective of the model training was to accurately predict the course load rating on a continuous five-point scale for all four constructs individually (i.e., time load, mental effort, psychological stress, and combined course load). In all of our experiments, we truncated all model predictions to the respective minimum and maximum scale ratings (i.e., 1 and 5). We evaluated models by mean absolute error ($MAE$).

We trained (1) a linear regression, (2) a random forest regressor, (3) an Elastic Net regressor, (4) a support vector machine regressor, (5) an XGBoost regressor, (6) a neural network regressor with ReLU activation function at the output and mean squared error loss, and (7) an ensemble prediction based on the mean of all predictors above. Model selection was based on the aim of having a variety of prediction schemes (both classical and contemporary) representing different architectural complexity and ensuring ease of reproducibility by using off-the-shelf libraries. An average rating prediction based on the training set served as the baseline for model performance. Reproducible code for model training and all data analysis code is in a public GitHub repository.\footnote{\url{https://github.com/CAHLR/courseload-analytics-LAK}} 

\subsection{Model evaluation}

At cross-validation, we compared three scale types (i.e., magnitude, manageability, and their average with manageability inverted) and two imputation strategies, resulting in a total of 216 model fits across our machine learning architectures. We compared a $k$-means imputation strategy ($k=2$) to a control variable imputation strategy as described in Section \ref{sec:method:lmsdata}. To select a final preprocessing strategy for test set evaluation, we computed the percentage improvement in $MAE$ compared to the average baseline. Scaling the difference in $MAE$ by the error of the average baseline takes into account the different spread (i.e., $SD$) across course load constructs in their ground-truth labels. During test set evaluation, we chose one model architecture and course load construct for retrospective model scaling based on $MAE$ after training our final selection of candidate models on all cross-validation data. We check whether the improvement in $MAE$ is statistically significant by computing confidence intervals for each $MAE$ via bootstrapping ($N = 1{,}000$; $\alpha = 0.05$). 

\subsection{Longitudinal model scaling and its analysis}
\label{sec:methods:model-scaling}

We gathered course load predictions for all courses in the enrollment records of UC Berkeley. Where LMS data for a particular semester is missing, we imputed via average course predictions across semesters with available LMS data. The rationale behind this is that fluctuation in instructional features encoded in LMS data (e.g., the engagement of students and instructors in the LMS forum) is expected to balance out across multiple course iterations, assuming that the course structure does not substantially change. We imputed course load predictions for 12.56\% of course offerings based on an average of 2.80 courses. Hence, for courses with no course load prediction for a specific semester, course load features from an average of 2.80 earlier or later course iterations was used for prediction. We defined course enrollments as courses students did not drop before the fifth week of registration, when students can make no more changes to enrollments. To have a fair comparison of semester course loads across semesters of enrollment, we restricted our sample to students who entered UC Berkeley in the Fall 2017 semester until nominal graduation in Spring 2021 ($N = 58{,}325$ student semester pairs with $N = 9{,}106$ students). We did this because program and course prerequisites can change based on a cohort-by-cohort basis. Given that transfer students who entered UC Berkeley in Fall 2017 nominally are in their fifth semester, we compared those students to their non-transfer peers from the Fall 2015 intake.

To investigate the longitudinal development of semester load, we summed up the credit hour designations and predicted course loads of courses taken by students each semester. We abbreviate the predicted semester course load as $PCL_{sem}$. To compare course load across semesters of enrollment across student groups, we aggregated the average course load in credit hours and predicted load for each semester. To investigate the association between semester load and academic outcomes (i.e., delayed graduation and degree stop-out), we employed logistic regression models. We compared models of academic outcomes based on credit hours to an additive and interaction model of both semester loads using likelihood-ratio tests. In a secondary analysis, we aggregated the ratio with which enrollments in a given course were by students who ended up leaving their program of study. We then calculated the course load discrepancy between predicted course load and designated credit hours for each course as the difference in $SDs$ between both distributions. Finally, we investigated correlates of that discrepancy via partial correlation, controlling for credit hours.

\section{Results}

We begin reporting the accuracy with which our machine learning models can predict course load. We then scale the best-performing model to historical enrollment data and report longitudinal differences in predicted semester load ($PCL_{sem}$) across students. We report associations between our prediction and academic outcomes, prompting scrutiny of discrepancies between credit hour designation and course load prediction at the course level.

\subsection{RQ1: Accuracy of course load prediction using LMS, enrollment, and course embedding features}

We used cross-validation to ascertain an imputation strategy and scale selection for test set evaluation, totaling 108 model fits. We lead with the test set prediction results, followed by observations at the cross-validation step. 

A detailed breakdown of the test set model performances is in Table \ref{tab:model-performance}. Combined course load exhibited the lowest $MAE$ at 0.626 scale points and largest \% improvement, with the ensemble model improving the $MAE$ by 21.9\% compared to baseline. Notably, time load exhibited the lowest $MAE$ improvements at 9.97\% for a linear regression model and an $MAE$ of 0.720 scale points. Next to combined course load, the ensemble model performed best overall, with an average rank of two across all four machine learning tasks. All average model ranks are in Table \ref{tab:model-rank}. 

\begin{table}[htp]
\caption{Holdout test set performance across machine learning tasks and models, including $p$-values indicating whether the improvement in mean absolute error ($MAE$) is significant ($p < 0.05)$; colored in green) or marginally significant ($p < 0.10$; colored in yellow) when compared to an average baseline error estimate ($N = 1{,}000$ bootstrap samples).}
\label{tab:model-performance}
\begin{tabular}{r|rrrrrrr}
\textbf{Outcome} & \cellcolor[HTML]{FFFFFF}\textbf{Model} & \cellcolor[HTML]{FFFFFF}\textbf{Rank} & \cellcolor[HTML]{FFFFFF}\textbf{$\pmb{MAE}$} & \cellcolor[HTML]{FFFFFF}\textbf{$\pmb{\Delta MAE}$} & \cellcolor[HTML]{FFFFFF}\textbf{\% Improve} & \cellcolor[HTML]{FFFFFF}\textbf{$\pmb{MAE}$ CI 95\%} & \cellcolor[HTML]{FFFFFF}\textbf{\textit{p}} \\
\hline
\rowcolor[HTML]{F5F5F5} 
Time Load             & Linear Regression                                 & 1                                            & 0.720                                & 0.080                                      & 9.97                                                              & [0.580, 0.876]                                 & .148                              \\
\rowcolor[HTML]{FFFFFF} 
            & Ensemble                               & 2                                            & 0.729                                & 0.071                                      & 8.87                                                              & [0.580, 0.887]                                  & .181                              \\
\rowcolor[HTML]{F5F5F5} 
             & SVM                                    & 3                                            & 0.730                                & 0.070                                      & 8.71                                                              & [0.533, 0.939]                                  & .247                              \\
\rowcolor[HTML]{FFFFFF} 
            & Neural Network                                     & 4                                            & 0.754                                & 0.045                                      & 5.64                                                              & [0.594, 0.923]                                 & .284                              \\
\rowcolor[HTML]{F5F5F5} 
             & Elastic Net                                   & 5                                            & 0.787                                & 0.013                                      & 1.60                                                              & [0.634, 0.951]                               & .435                              \\
\rowcolor[HTML]{FFFFFF} 
             & Random Forest                                     & 6                                            & 0.797                                & 0.002                                      & 0.28                                                              & [0.641, 0.959]                              & .472                              \\
\rowcolor[HTML]{F5F5F5} 
             & XGBoost                                    & 7                                            & 0.799                                & 0.000                                      & 0.06                                                              & [0.650, 0.956]
             & .486                              \\
\rowcolor[HTML]{FFFFFF} 
            & Average Baseline                                 & 8                                            & 0.799                                & 0.000                                      & 0.00                                                              & [0.642, 0.962]                 & .488                              \\
\hline
\rowcolor[HTML]{93C47D} 
Mental Effort              & XGBoost                                    & 1                                            & 0.737                                & 0.164                                      & 18.21                                                             & [0.578, 0.903]                   & .026                              \\
\rowcolor[HTML]{93C47D} 
              & Ensemble                               & 2                                            & 0.746                                & 0.155                                      & 17.24                                                             & [0.592, 0.911]                             & .033                              \\
\rowcolor[HTML]{FFF2CC} 
              & Elastic Net                                   & 3                                            & 0.770                                & 0.131                                      & 14.53                                                             & [0.619, 0.934]                               & .058                              \\
\rowcolor[HTML]{FFF2CC} 
              & Linear Regression                                 & 4                                            & 0.780                                & 0.121                                      & 13.47                                                             & [0.625, 0.947]                                & .072                              \\
\rowcolor[HTML]{F5F5F5} 
              & Random Forest                                     & 5                                            & 0.819                                & 0.082                                      & 9.14                                                              & [0.647, 1.004]                               & .177                              \\
\rowcolor[HTML]{FFFFFF} 
              & Neural Network                                     & 6                                            & 0.874                                & 0.027                                      & 3.01                                                              & [0.666, 1.087]                            & .399                              \\
\rowcolor[HTML]{F5F5F5} 
              & Average Baseline                                 & 7                                            & 0.901                                & 0.000                                      & 0.00                                                              & [0.702, 1.106]                               & .492                              \\
\rowcolor[HTML]{FFFFFF} 
              & SVM                                    & 8                                            & 0.907                                & -0.006                                     & -0.67                                                             & [0.707, 1.127]                             & .504                              \\
\hline
\rowcolor[HTML]{FFF2CC} 
Psychological Stress              & XGBoost                                    & 1                                            & 0.962                                & 0.130                                      & 11.93                                                             & [0.800, 1.128]                              & .060                              \\
\rowcolor[HTML]{FFF2CC} 
              & Neural Network                                     & 2                                            & 0.980                                & 0.112                                      & 10.22                                                             & [0.816, 1.153]                                 & .097                              \\
\rowcolor[HTML]{F5F5F5} 
              & Ensemble                               & 3                                            & 0.991                                & 0.101                                      & 9.29                                                              & [0.832, 1.157]                               & .108                              \\
\rowcolor[HTML]{FFFFFF} 
              & Linear Regression                                 & 4                                            & 0.993                                & 0.099                                      & 9.07                                                              & [0.828, 1.167]                                & .130                              \\
\rowcolor[HTML]{F5F5F5} 
              & Elastic Net                                   & 5                                            & 0.997                                & 0.095                                      & 8.74                                                              & [0.822, 1.186]                              & .155                              \\
\rowcolor[HTML]{FFFFFF} 
              & Random Forest                                     & 6                                            & 1.036                                & 0.056                                      & 5.09                                                              & [0.865, 1.208]                             & .268                              \\
\rowcolor[HTML]{F5F5F5} 
              & SVM                                    & 7                                            & 1.059                                & 0.033                                      & 3.00                                                             & [0.885, 1.237]                            & .350                              \\
\rowcolor[HTML]{FFFFFF} 
              & Average Baseline                                 & 8                                            & 1.092                                & 0.000                                      & 0.00                                                              & [0.923, 1.266]                            & .496                              \\
\hline
\rowcolor[HTML]{93C47D} 
Combined Course Load    & Ensemble                               & 1                                            & 0.626                                & 0.176                                      & 21.92                                                             & [0.503, 0.757]                  & .004                              \\
\rowcolor[HTML]{93C47D} 
    & Elastic Net                                   & 2                                            & 0.635                                & 0.167                                      & 20.84                                                             & [0.505, 0.770]   & .007                              \\
\rowcolor[HTML]{93C47D} 
    & XGBoost                                    & 3                                            & 0.662                                & 0.140                                      & 17.41                                                             & [0.522, 0.815] & .035                              \\
\rowcolor[HTML]{93C47D} 
    & Neural Network                                     & 4                                            & 0.683                                & 0.119                                      & 14.78                                                             & [0.555, 0.821] & .046                              \\
\rowcolor[HTML]{FFF2CC} 
    & Linear Regression                                 & 5                                            & 0.707                                & 0.094                                      & 11.77                                                             & [0.589, 0.835] & .070                              \\
\rowcolor[HTML]{FFF2CC} 
    & Random Forest                                     & 6                                            & 0.718                                & 0.084                                      & 10.43                                                             & [0.593, 0.845] & .095                              \\
\rowcolor[HTML]{F5F5F5} 
    & SVM                                    & 7                                            & 0.740                                & 0.062                                      & 7.71                                                              & [0.569, 0.928] & .242                              \\
\rowcolor[HTML]{FFFFFF} 
    & Average Baseline                                 & 8                                            & 0.802                                & 0.000                                      & 0.00                                                              & [0.660, 0.948] & .489                             
\end{tabular}
\end{table}

\begin{table}[htp]
\caption{Average model rank of different model architectures on the holdout test set across the four machine learning tasks (i.e., time load, mental effort, psychological stress, combined course load).}
\label{tab:model-rank}
\begin{tabular}{l|rrrrrrrr}
Model & Ensemble & XGBoost  & Lin. Regression & Elastic Net & Neural Net   & Random Forest   & SVM  & Baseline \\
\hline
Avg Rank & 2.00     & 3.00 & 3.50   & 3.75 & 4.00 & 5.75 & 6.25 & 7.75  
\end{tabular}
\end{table}

At cross-validation, we find the best percentage improvement in cross-validation error for predicting our magnitude survey scales. Note that selection based on $MAE$ alone across different scales might not be a fair comparison given that the variance of these scales, and therefore their $MAE$ at mean prediction baseline, differ. Second, we find that imputing based on control variables lead to superior cross-validation $MAE$ for 72 out of 108 model fits, a frequency significantly higher than chance based on a simple binomial test ($p < .001$). Therefore, we performed our final test set evaluation on continuous measures of course load magnitude with a control variable imputing strategy.

\subsection{RQ2: Comparing course load throughout students' academic degree across STEM and transfer students}

\begin{figure}[htp]
  \begin{minipage}{\textwidth}
    \centering
    \includegraphics[width=.48\textwidth]{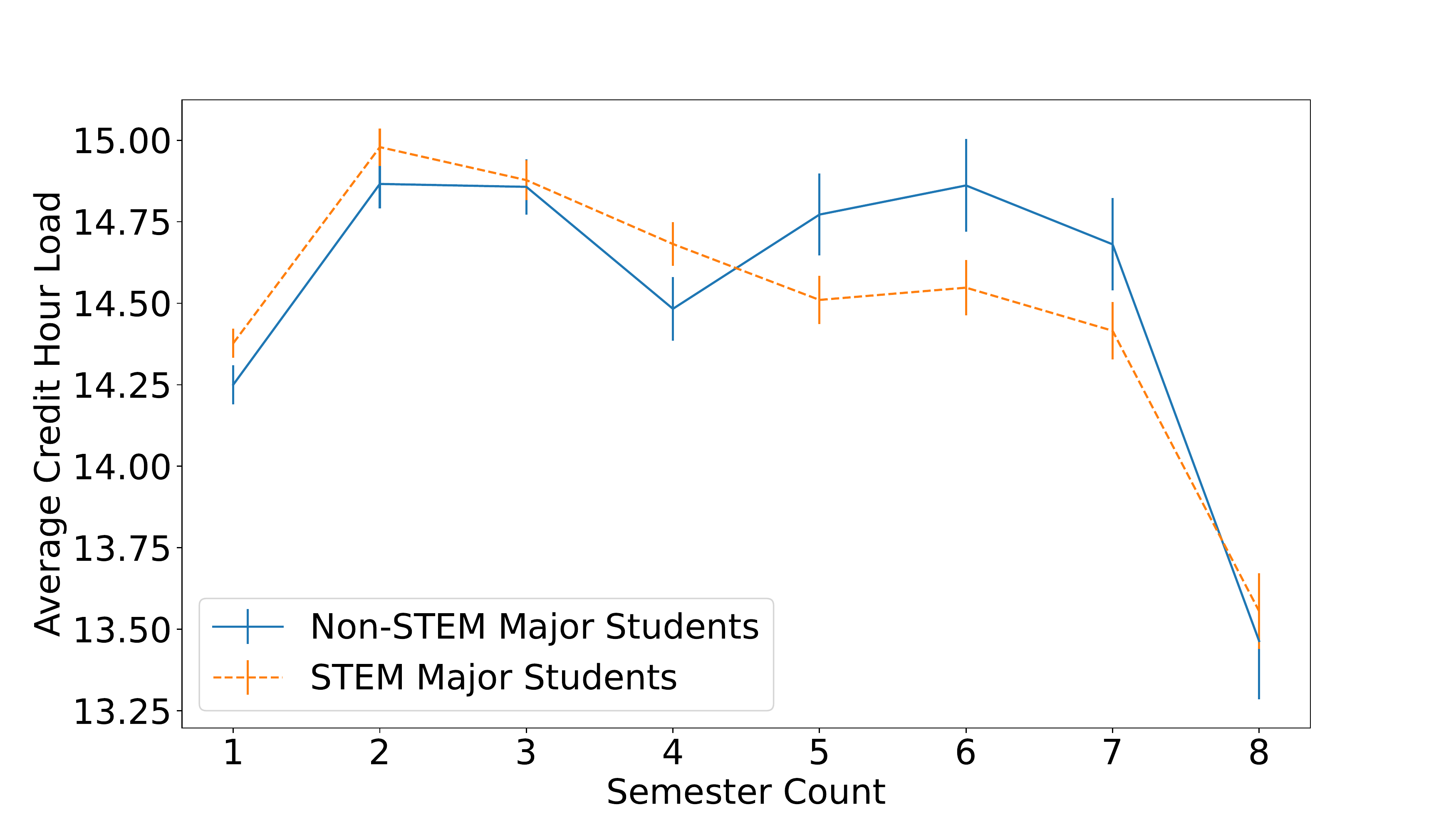}\quad
    \includegraphics[width=.48\textwidth]{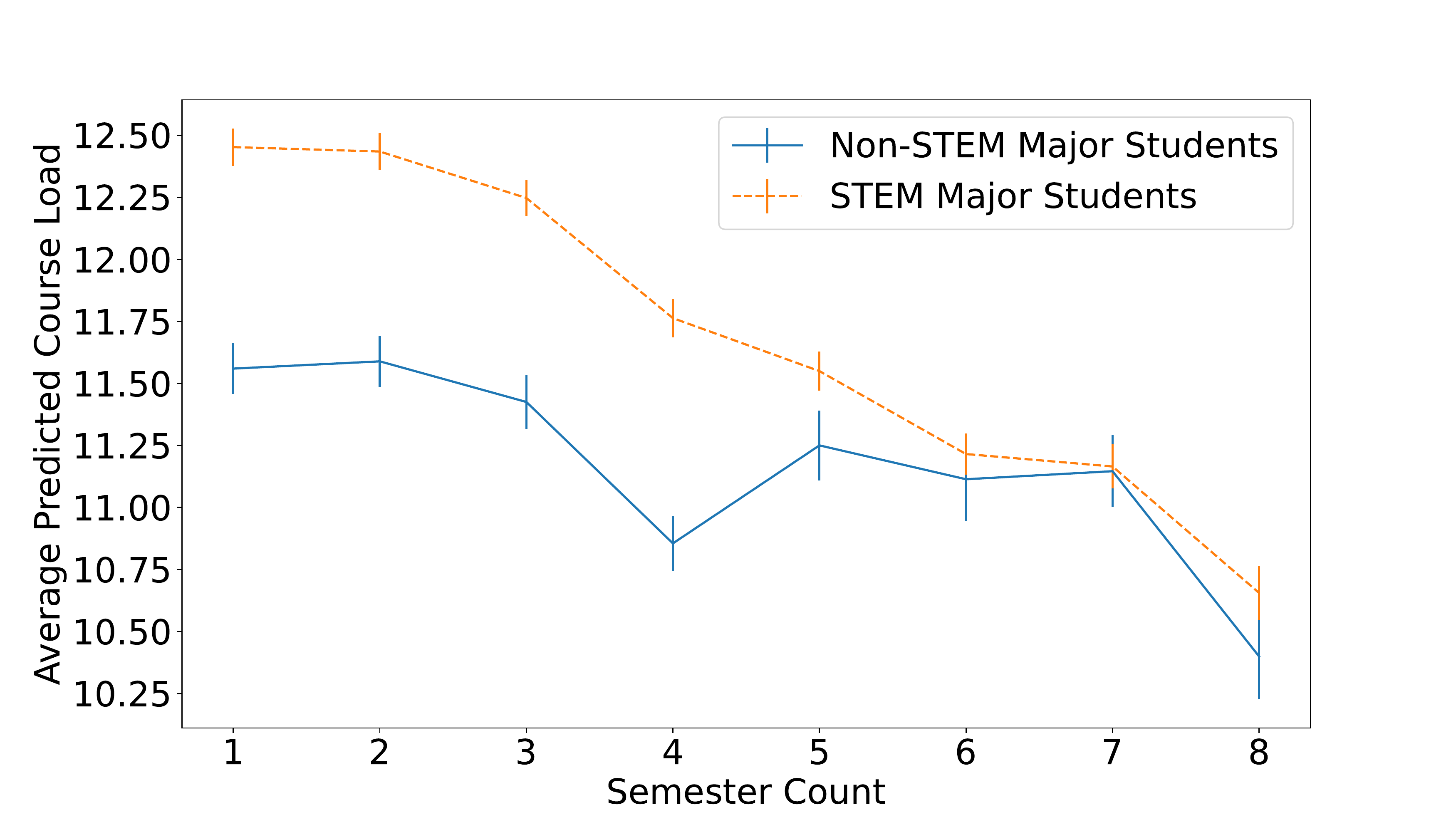}\\
  \end{minipage}
  \caption{Average credit hour semester load (left) and $PCL_{sem}$ (right) by semester of enrollment broken out by STEM students and non-STEM students, including two standard error bars.}
  \label{fig:stemcomp}
\end{figure}

RQ2 pertains to the longitudinal development of semester course load of STEM compared to non-STEM and transfer compared to non-transfer students. We compared students across credit hour semester load and $PCL_{sem}$ based on the model selected in RQ1. We start by comparing STEM to non-STEM students in Figure \ref{fig:stemcomp}.

\begin{figure}[htp]
  \begin{minipage}{\textwidth}
    \centering
    \includegraphics[width=.48\textwidth]{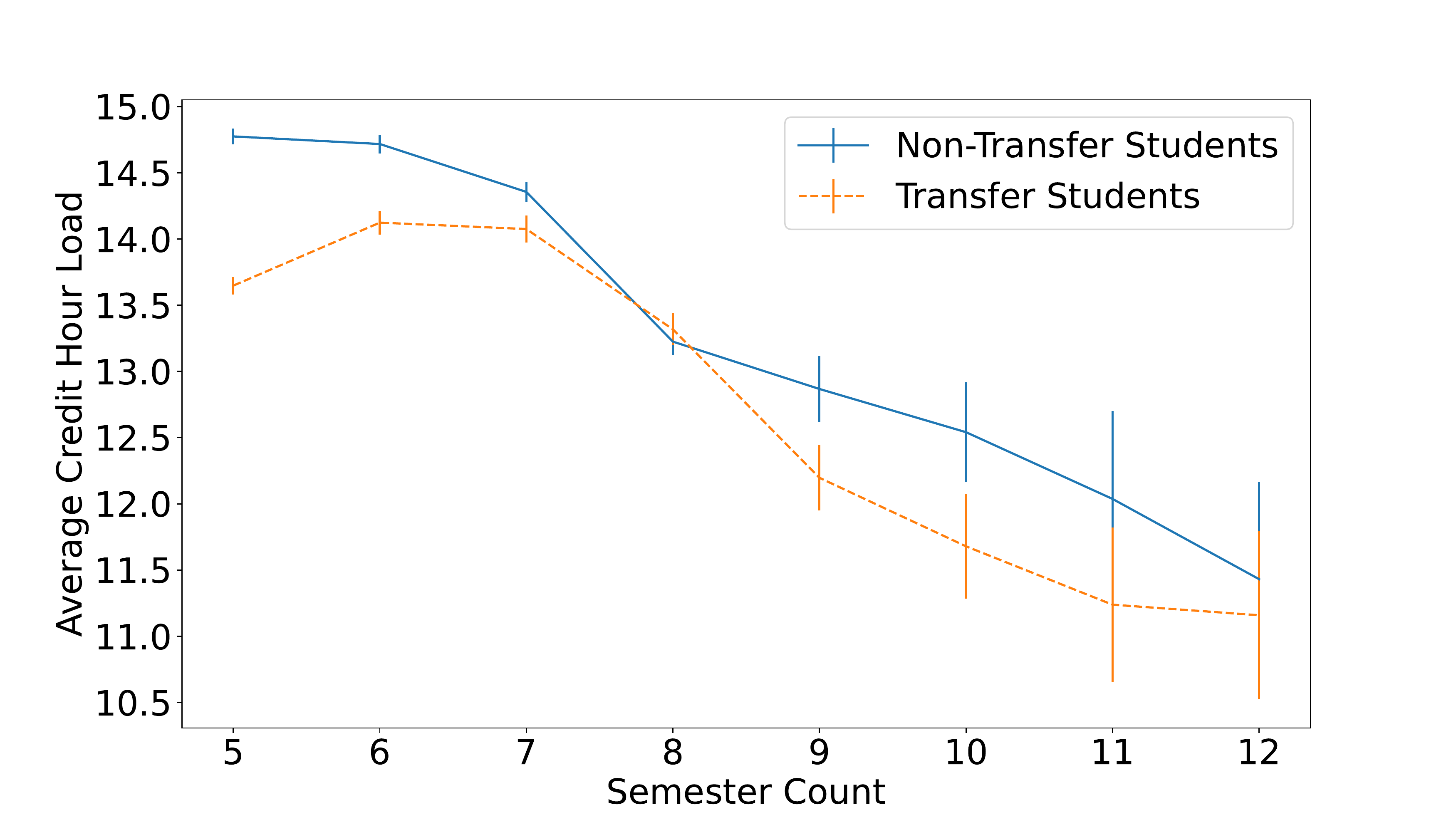}\quad
    \includegraphics[width=.48\textwidth]{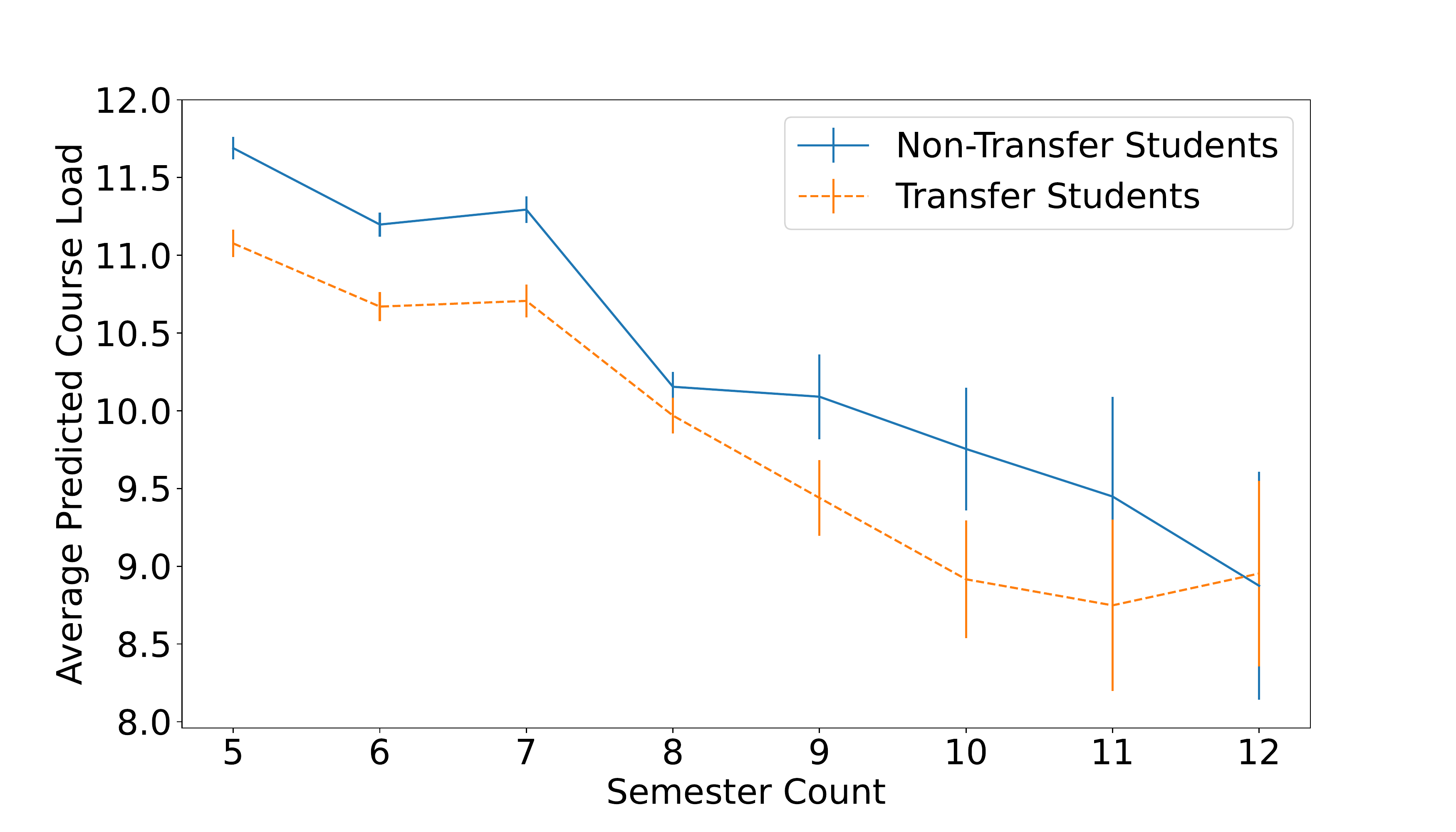}\\
  \end{minipage}
  \caption{Average credit hour semester load (left) and $PCL_{sem}$ (right) by semester of enrollment broken out by transfer and non-transfer students of the same cohort, including two standard error bars.}
  \label{fig:transfercomp}
\end{figure}

We find that CLA predicted a significantly larger mean semester load in STEM compared to non-STEM students in their first four semesters of enrollment. Notably, while credit hour load dropped by around 3\% for STEM and non-STEM students from semesters two to four, $PCL_{sem}$ indicated that this drop is around 6\% for STEM students and 4\% for non-STEM students. Notably, the first semester was among the highest in course load according to $PCL_{sem}$ while it was among the lowest according to credit hour semester load. We report analogous plots comparing transfer students entering UC Berkeley in Fall 2017 with their peers in their fifth semester in Fall 2017 in Figure \ref{fig:transfercomp}.

Credit hour semester loads suggested that the gap between transfer and non-transfer students closed in the first four semesters of enrollment. CLA indicated that the difference in load between transfer and non-transfer students was constant across the first four semesters, with semester-to-semester trends following a visibly parallel trajectory.

\subsection{RQ3: Associations of predicted and credit hour load with stop-out and on-time graduation}

\subsubsection{Stop-out and on-time graduation}

\begin{figure}[htp]
  \begin{minipage}{\textwidth}
    \centering
    \includegraphics[width=.48\textwidth]{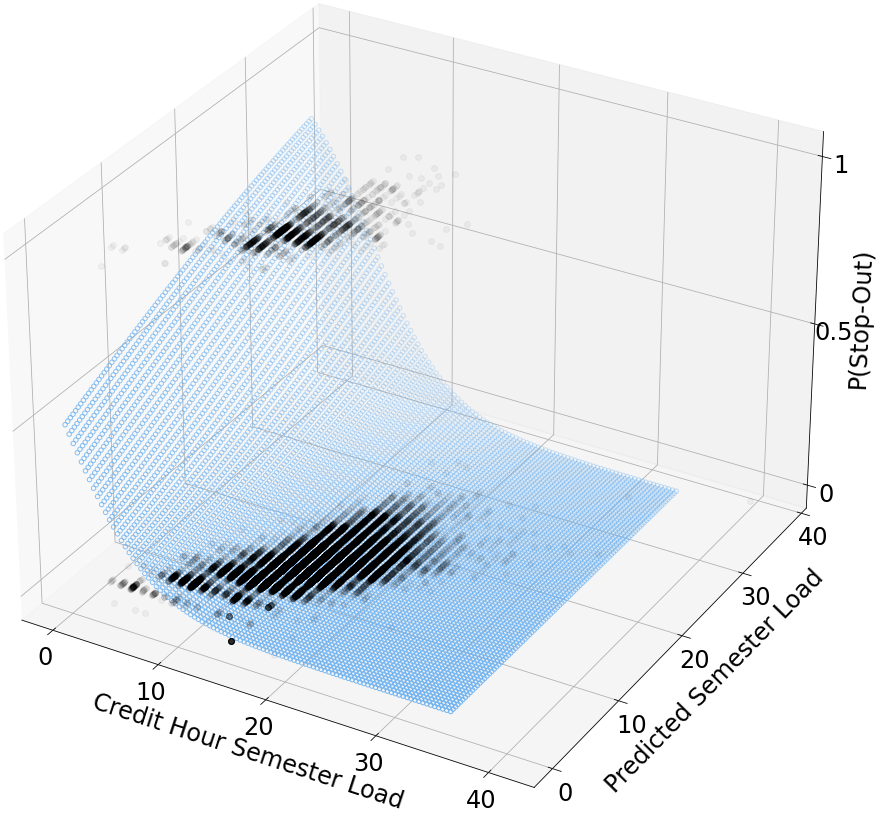}\quad
    \includegraphics[width=.48\textwidth]{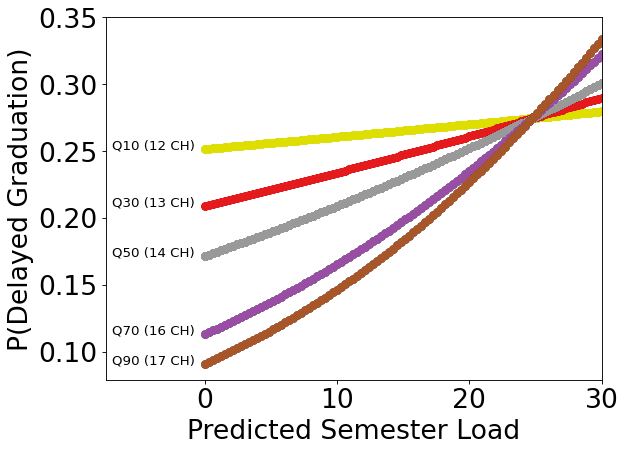}\\
  \end{minipage}
  \caption{Inferred marginal effect of $PCL_{sem}$ on the probabilities of stop-out, including empirical observations in black (left) and delayed graduation when controlling for credit hours (CH) at different quantiles (Q) of their distribution (right).}
  \label{fig:modelspace}
\end{figure}

We tested linear binomial models of whether students ended up leaving their program of study or delaying their graduation given their semester credit hours and $PCL_{sem}$. We compared a model featuring credit hour semester load to a model with additive effects of $PCL_{sem}$ and a model with an interaction of both loads. For degree stop-out, based on a likelihood-ratio test, we find that the additive model fit the data best, outperforming the credit hour model ($\chi^2(1) = 29.22, p < .001$). Employing McFadden's $R^2$ metric for logistic regression, we find that adding $PCL_{sem}$ approximately doubled explained variance from $R^2$ = 0.21\% to $R^2$ = 0.39\%. In contrast, the interaction model did not increase fit ($\chi^2(1) = 0.43, p = .512$, $R^2$ = 0.39\%). For on-time graduation, a model with an interaction of credit hour and predicted load fit best, outperforming the additive model ($\chi^2(1) = 130.40, p < .001$, $R^2$ = 1.70\%) while credit hours alone explained $R^2$ = 1.40\%. We plot model inferences across semester credit hour loads in Figure \ref{fig:modelspace} to understand the learned model parameters of our selected models.

We find that our selected model of stop-out inference predicted that students with a low credit hour but comparatively high $PCL_{sem}$ have a higher probability of leaving their program of study. For delayed graduation, $PCL_{sem}$ was positively associated with graduation delay for high and unrelated with graduation delay for low credit hour semester load. Notably, there seems to be a crossover point for delayed graduation where all credit hour semester loads are inferred to have a delayed graduation probability of around 27.5\%.

Based on these findings, we asked whether our model inferences are compatible with UC Berkeley's policy of requiring students to obtain special permission to take more than 20 semester credits in its largest college.\footnote{\url{https://lsadvising.berkeley.edu/policies/unit-minimum-maximum-semester}} Given that the univariate association between credit hour semester load (i.e., ignoring predicted load) and delayed graduation was negative and did not suggest a cutoff, we mapped our course load analytic onto credit hours to answer that question. Three credit hour courses are the most common courses offered at UC Berkeley. Therefore, we took the median predicted load for a three credit hour course (2.62) and estimated, based on the crossover point on the right panel of Figure \ref{fig:modelspace}, that 28.62 credit hour equivalents would be associated with a delayed graduation probability of around 27.5\%, that is, an increased probability compared to the average baseline graduation delay probability (21.55\%). This cutoff for increased delayed graduation risk would be higher than the institutional guidelines of 20 credit hours. However, as our findings suggest, this cutoff can be tailored to different students' work packages as selected by them a priori.

The stop-out model exhibited an $AUC$ of 0.57 ($CI_{95\%} = [0.54, 0.61]$) and the delayed graduation model exhibited an $AUC$ of 0.55 ($CI_{95\%} = [0.54, 0.56]$) which is not satisfactory. However, the findings motivate an analysis of course load discrepancy, which we defined as the difference between credit hour and predicted load after standardizing both measures to a mean of 0 and a standard deviation of 1. We focused our analyses of course load discrepancy on stop-out, given the relatively large contribution of $PCL_{sem}$ to stop-out compared to on-time graduation.

\subsubsection{Exploratory analyses of predicted course load discrepancy}

We calculated a workload discrepancy metric between standardized credit hours and predicted load. To avoid the discrepancy index being more unreliable in more extreme ranges (e.g., high positive discrepancy), we checked the association between the mean absolute error (our model evaluation metric) and the discrepancy index in our training and test data. We find that model accuracy was uncorrelated with discrepancy to credit hours for the full rated course load sample ($r(330) = 0.04, p = .500$) as well as in our out-of-sample holdout test set ($r(48) = -0.18, p = .230$). Given course load prediction differences between STEM and non-STEM in Figure \ref{fig:stemcomp}, we investigated the distribution of course load discrepancy within STEM and non-STEM courses.

We find that STEM courses had a significantly higher positive discrepancy ($M = 0.54, CI_{95\%} = [0.47, 0.61]$) than non-STEM courses ($M = -0.35, CI_{95\%} = [-0.40, -0.30]$), $p < .001$. Employing further two-sided $t$-tests, we find that for courses with a high ratio of students enrolled that left their program, the discrepancy index was higher within STEM ($M = 0.91, CI_{95\%} = [0.53, 1.39]$ compared to $M = 0.52, CI_{95\%} = [0.45, 0.59]$, $p = .002$), but not within non-STEM courses ($M = -0.33, CI_{95\%} = [-0.47, -0.19]$ compared to $M = -0.35, CI_{95\%} = [-0.41, -0.30]$, $p = .763$).

What are the most contributing features to course load discrepancy? We calculated the correlation between our features and course load discrepancy to credit hours for our chosen ensemble model and all courses ($N = 4{,}305$). We controlled for credit hours and regression to the mean via partial correlation. Table \ref{tab:discrepancy-determinants} shows the top 15 correlations.

\newcommand{\boldm}[1] {\mathversion{bold}#1\mathversion{normal}}
\begin{table}[htp]
\caption{Top 15 most correlated features (excluding course2vec) with course load discrepancy when controlling for credit hours via partial correlation, including 95\% confidence intervals obtained via bootstrapping ($N = 1{,}000$).}
\label{tab:discrepancy-determinants}
\begin{tabular}{rrrr}
\rowcolor[HTML]{FFFFFF} 
\cellcolor[HTML]{FFFFFF}\textbf{Feature}          & \boldm{$r_{partial}$} & \textbf{\textit{CI 95\%}}   & \textbf{\textit{p}}      \\
\rowcolor[HTML]{F5F5F5} 
N Satisfied Prereqs (All Past Semesters)       & 0.54           & {[}0.51, 0.58{]}   & \textless{}.001 \\
\rowcolor[HTML]{FFFFFF} 
\% Satisfied Prereqs (All Past Semesters) & 0.52           & {[}0.49, 0.56{]}   & \textless{}.001 \\
\rowcolor[HTML]{F5F5F5} 
Course Grade Standard Deviation                         & 0.27           & {[}0.22, 0.31{]}   & \textless{}.001 \\
\rowcolor[HTML]{FFFFFF} 
N Satisfied Prereqs (Current Semester)         & 0.21           & {[}0.16, 0.25{]}   & \textless{}.001 \\
\rowcolor[HTML]{F5F5F5} 
Course is a STEM Course                              & 0.19           & {[}0.16, 0.22{]}   & \textless{}.001 \\
\rowcolor[HTML]{FFFFFF} 
\% Satisfied Prereqs (Current Semester)    & 0.18           & {[}0.13, 0.23{]}   & \textless{}.001 \\
\rowcolor[HTML]{F5F5F5} 
Student is a STEM Student                                 & 0.14           & {[}0.08, 0.20{]}    & \textless{}.001 \\
\rowcolor[HTML]{FFFFFF} 
Number of Enrolled Teaching Staff (LMS)                      & 0.14           & {[}0.11, 0.17{]}   & \textless{}.001 \\
\rowcolor[HTML]{F5F5F5} 
Total Number of Listed Prereqs                                        & 0.13           & {[}0.10, 0.17{]}   & \textless{}.001 \\
\rowcolor[HTML]{FFFFFF} 
\% Percentage of Non-Letter Grades               & 0.13           & {[}0.08, 0.17{]}   & \textless{}.001 \\
\rowcolor[HTML]{F5F5F5} 
Instructional Staff Forum Reply Time to LMS Dropout                 & 0.11           & {[}-0.02, 0.23{]}  & 0.086           \\
\rowcolor[HTML]{FFFFFF} 
Students did not Use LMS Forum                                 & 0.10           & {[}0.07, 0.13{]} & \textless{}.001 \\
\rowcolor[HTML]{F5F5F5} 
Student GPA                                     & -0.10           & {[}-0.13, -0.07{]} & \textless{}.001 \\
\rowcolor[HTML]{FFFFFF} 
Student Major GPA                               & -0.11           & {[}-0.14, -0.08{]} & \textless{}.001 \\
\rowcolor[HTML]{F5F5F5} 
Course GPA                       & -0.25           & {[}-0.30, -0.21{]} & \textless{}.001
\end{tabular}
\end{table}

Based on Table \ref{tab:discrepancy-determinants}, we find that most discrepancy determinants related to prerequisites. The number and percentage of satisfied prerequisites in all past semesters ($r = 0.54$ and $r = 0.52$, respectively) were most strongly and positively associated with course load discrepancy, which was surprisingly higher than the association of the total number of listed prerequisites with course load discrepancy ($r = 0.13$). Notably, the observed course load discrepancy was also positively associated with the STEM status of the course and student ($r = 0.19$ and $r = 0.14$, respectively). 

How robust are these correlations for the subsample of courses with more than two $SDs$ discrepancy to designated credit hour load ($N = 388$)? Five out of 15 top features out of the full sample were also significant for the subsample of courses with high positive discrepancy. These include the frequency of satisfied prerequisites (in all past and current semesters) and the total number of prerequisites. Taken together, the features that determine discrepancy primarily comprised signal from course prerequisites. This finding is robust for highly discrepant courses in our sample.

The observation that prerequisite-based features were associated with course load discrepancy also showed in our course2vec features. We fit two multiple linear regression models to infer course load discrepancy through these features while controlling for the number of credit hours. In particular, we compared the explained variance in course load discrepancy by the course2vec features and the average course2vec features of the course's prerequisites. We find that the average prerequisite vector model explained significantly more variance in course load discrepancy ($R^2 = 0.75, CI_{95\%} = [0.71, 0.80]$) than the model including the vector of the course itself ($R^2 = 0.63, CI_{95\%} = [0.61, 0.65]$). For reference, the variance in course load discrepancy explained by credit hours amounted to $R^2 = 0.50, CI_{95\%} =[0.47, 0.53]$. 

How much adjustment in credit hour designations would courses with high positive course load discrepancy (i.e., more than two $SDs$) need? We fit a regression of predicted load over the number of satisfied prerequisites and find that this model adds $\beta=0.59$ predicted units per satisfied prerequisite. Based on this, we looked at the average number of satisfied prerequisites across students in these courses and converted the increase in predicted load back to credit hours based on our earlier formula (three credit hour courses equalling 2.62 $PCL$ units). An average of 0.37 credit hours adjustment would be necessary to accommodate the discrepancy in these courses. This adjustment was about 0.40 units for courses with a discrepancy of three $SDs$ and remains relatively constant based on different filtering thresholds.  

What qualitative observations can be made about the courses with an exceptionally high (positive) discrepancy between credit hour designation and $PCL$ based on our analytics? We manually investigated the features and characteristics of courses with more than four $SDs$ differences between the standardized distributions of credit hours and predicted course load ($N = 31$). We find that 21 out of 31 had at least ten listed prerequisites (which may be listed as alternatives). In addition, these courses had an average student-to-instructor ratio of $M=12.87$ while this quantity was $M=20.62$ for all courses in our sample. Seven of the 31 courses had more than one graded weekly assignment. Regarding the distribution of subjects and majors, while 17 out of 31 courses were STEM courses, we find that seven, (i.e., 50\% of non-STEM courses) were music classes. These courses included performance-based courses ($N=2$), courses on harmony ($N=3$), courses on culture and history ($N=1$), and singing and hearing training ($N=1$).

\section{Discussion}

We find that course load was most accurately predicted for a combined construct of time load, mental effort, and psychological stress regarding $MAE$ improvement and overall $MAE$. As our features picked up more signal in mental effort than in time load and psychological stress, our selected model may overrepresent mental effort. Future work is required to improve prediction in time load and psychological stress while potentially improving upon our scale reliability estimates by developing more survey questions capturing facets of time load and psychological stress. This improvement may be within reach, as several relevant features can be identified that have been unexplored concerning course load due to their current inaccessibility for our study. One may quantify time load via assignment or essay contents (e.g., text complexity to gauge assignment effort) \cite{jackson2016common}. We purposely omitted the content of assignments and submissions from our dataset to preserve privacy. Content analysis methods from learning analytics \cite{rose2017discourse} may also be applied to other artifacts, for example, LMS discussion and Zoom chats, to assess course load. Furthermore, competitive course grading on a curve \cite{marks2007unsettled}, or sense of belonging resulting from a match in demographic attributes between instructor and student \cite{cheryan2020double} may relate to psychological stress. 

Our analytic suggests that course load may be particularly high in students' first semester, which is not reflected in overall credit hour semester load. This discrepancy in loads has potential implications for how institutions handle the experience of first-year university students. Second-year college retention and high school preparation have been a focus of prior work on stop-out and dropout \cite{herzog2005measuring}. Future research may investigate whether reducing or modifying first-semester requirements according to CLA improves retention. We find prerequisites, particularly the number of satisfied prerequisites, to be positively associated with load discrepancy to credit hours. This may be due to a selection bias, where students at UC Berkeley that bring more AP credit have fewer satisfied prerequisites when enrolling in courses. If it is true that students with less preparation that come to UC Berkeley take more prerequisites, future research may test whether interventions targeting implied high school prerequisites in first-year students improves student retention. Supplementary courses may be served via intelligent tutoring systems during summer, potentially in hybrid settings, which prior work found time-efficient concerning learning gains \cite{lovett2008open}.

We gain novel insights regarding stop-out and on-time graduation by scaling CLA to historical enrollment, which contributes to understanding both phenomena. For stop-out, our inferential model suggests that students with a relatively low credit hour load but high load according to our metric are more likely to leave their degree. We find STEM courses with exceptionally high predicted load with respect to their credit hour designation associated with a high frequency of stop-out enrollments. For these courses, our analytic suggests adding 0.40 credit hours of adjustment. These discrepancies were chiefly related to course prerequisites and their satisfaction. This finding aligns with prior works describing math preparedness as a correlate of STEM degree attainment \cite{chen2013stem, cohen2020mathematics, zhang2022early}. 

Among students who graduate, we find that predicted course load becomes increasingly important for inferring on-time graduation the higher the semester credit hour load is. We interpret this as students who underenroll on credit not being able to satisfy degree requirements on time, irrespective of the load they incur according to our metric. Another potential reason for diminishing importance of predicted load is that students who underenroll may compensate better for work-intensive courses as they have more excess resources (e.g., leisure time) to accommodate high course load as argued in \cite{huntington2021semester}. Finally, our models suggest a crossover point in predicted load irrespective of credit hours at which there is an increased risk of delayed graduation. We estimated this load to be 28.62 semester units, which is more than UC Berkeley designates as the upper bound for normative enrollment. For real-world deployment of CLA, stop-out may be a more fruitful avenue than on-time graduation, as explained variance was comparatively low for the latter.

CLA has implications for various stakeholders in higher education. Students may benefit from CLA by either consuming CLA as an additional point of information when planning their semester or in a recommendation system setting in which course selection options can be constrained based on a desired course load package. These recommendations could also consider students' prior course history, as previous work indicates that a mismatch between students' current goal achievement and assignment difficulty may result in a vicious cycle of procrastination \cite{waschle2014procrastination}. Therefore, future research may leverage longitudinal student features at prediction, such as students' past success in managing course load or completing demanding courses and assignments. CLA may also help academic advisors understand students' course load challenges currently not reflected in credit hours. Given our findings regarding the discrepancy of CLA and credit hours, CLA may help remedy past findings reporting both underenrollment and overenrollment to be negatively associated with academic outcomes \cite{boumi2021quantifying,karimi2022causal}. Perhaps, instructors could cooperate with an algorithm flagging excess course load compared to credit hours during advising. Additionally, institutions and faculty may use CLA for designing and improving academic programs. Encouraging findings exist for time-related metrics gleaned from program compositions to enhance their curriculum \cite{ochoa2016simple}. As institutions may project CLA onto longitudinal undergraduate pathways given different course compositions, CLA may help assign workload across semesters in a program of study according to the desired distribution. Similarly, CLA may help detect inequalities among different options to fulfill program requirements. Different workloads among program requirement satisfaction options may result in imbalances in student enrollment numbers and, perhaps, in dissimilar learning gains, leading to unequal skill levels within cohorts that threaten program quality standards. Finally, CLA may help instructors reflect on their courses' workload regarding instructional course features, potentially in an evaluative monitoring and feedback tool used throughout the semester.

\subsection{Limitations and future work}

We recognize two limitations to this study that may guide future work. First, our data can not tell whether an adjustment of credit hour designations will lead to more desirable student outcomes. It is to be determined whether having students switch their semester course selection based on our analytics will change educational outcomes in the way our inferential models and analyses propose. Relatedly, future research may investigate whether students knowing course load discrepancies ahead of time leads to mindset adjustment or improved preparation for these courses. Students' course selection, particularly regarding semester credit hour loads, may reflect self-selection (e.g., students with high ability choosing particular course sets), and little work exists on student course selection strategies \cite{chaturapruek2021studying}. Therefore, future work may scrutinize students' assumptions when choosing courses based on CLA. For example, students may try to select courses while staying below a certain load threshold to avoid having to drop a course later in the semester. We note that these assumptions may systematically differ across demographic characteristics, calling for replication studies of our findings to other institutions to gauge generalizability and ensure equitable translation of CLA into effective advising strategies. Future work may investigate whether individualizing CLA to demographic or academic attributes (e.g., open degree requirements) leads to more desirable predictive properties. In addition, as factors other than workload may play into student course selection (e.g., employment prospects, taking courses with friends), it is yet to be investigated whether further validation of CLA with academic outcomes is desirable and feasible. Perhaps, predicting student course load ratings suffices for effective student-facing applications.

Second, our characterization of semester load is quite simplistic, taking the sum or predicted load for each course in a given semester. Future work might discover interactions among a particular set of courses by tracing course load on a more granular level, perhaps, weeks. Prior work suggests that time spent on courses increases before assignment deadlines \cite{ruiz2011assessing}. Therefore, overlapping spikes in workload among a particular course selection set may predict course drops or other academic outcomes. Future research may explore how the within-course workload composition given specific offerings relates to student outcomes and how overlapping spikes in workload across courses can be mitigated, for example, by asking for instructor deadline accommodations ahead of time. In addition, further ablation studies to investigate the importance of such instructional features could be compared to subsets of our current feature space.

\section{Summary and conclusions}

We have established course load analytics, a novel methodology for predicting higher education course workload from a variety of institutional data sources. An ensemble of machine learning models established a state-of-the-art 22\% improvement over an average baseline in predicting combined course load, an aggregation of time load, mental effort, and psychological stress ratings. With half a survey scale point in average out-of-sample accuracy, our model verges on readiness for reasonable deployment in student-facing applications. We retrospectively predicted course load patterns throughout students' four-year undergraduate experience by scaling our model to historical enrollment data. We found that a student's first semester at the University was their highest or second highest load semester as predicted by CLA, but the second lowest when calculated in terms of credit hours. In line with this finding, inferential models indicated that students with a low credit-hour semester load but comparatively high predicted load are more prone to degree stop-out. Differences between credit hour designation and predicted load were predicated on course prerequisites and particularly associated with STEM stop-out. Our analysis suggests that institutions may adjust courses with high load discrepancy to receive an additional 0.40 credit hours. Given the established capability to produce CLA and the role workload appears to play in student outcomes, we believe CLA is a fruitful avenue for future real-world experiments with student course selection in higher education and the continued study of undergraduate pathways.

\begin{acks}
We thank the reviewers for their helpful feedback to strengthen our manuscript. We thank UC Berkeley Research, Teaching, and Learning for their efforts with LMS data retrieval supported by the EVCP. Funding to attend this conference was provided by the CMU GSA/Provost Conference Funding and Schmidt Futures. 
\end{acks}

\bibliographystyle{ACM-Reference-Format}
\bibliography{main}

\end{document}